\documentclass[aps,pra,showpacs,twocolumn]{revtex4}
\usepackage{amsmath,graphicx,subfigure}

\def\q#1{$|#1\rangle$}
\def\qq#1#2{$|#1\rangle_{#2}$}
\def\ket#1{\mathinner{|{#1}\rangle}}

\begin{document}
\title{Quantum computing with an inhomogeneously broadened ensemble of ions: 
Suppression of errors from detuning variations by 
specially adapted pulses and coherent population trapping.}
\author{Ingela Roos$^{1}$ and Klaus M{\o}lmer$^{2}$}
\affiliation{$^{1}$ Department of Physics, Lund Institute of Technology, Box 118, S-211 00 Lund, Sweden \\
$^{2}$ QUANTOP, Danish National Research Foundation Center for Quantum Optics \\
Department of Physics and Astronomy, University of Aarhus, DK-8000 {\AA}rhus C, Denmark}

\begin{abstract}
The proposal for quantum computing with rare-earth-ion qubits
in inorganic crystals makes use of the inhomogeneous 
broadening of optical transitions in the ions to associate individual qubits with
ions responding to radiation in selected frequency channels. We show that a class of
Gaussian composite pulses and complex sech pulses provide accurate qubit $\pi$-rotations,
which are at the same time channel selective on a 5 MHz frequency scale and tolerant
to $\pm 0.5$ MHz deviations of the transition frequency of ions within a single channel.  
Rotations in qubit space of arbitrary angles and phases are produced by sequences of
$\pi$-pulses between the excited state of the ions and coherent superpositions of the 
qubit states. 
\end{abstract}
\maketitle

\section{introduction}
Rare-earth ions doped into inorganic crystals
experience inhomogeneous broadening. If the doping
concentration is high, a near field probe is able to
address individual ions and to induce resonant excitation hopping among nearby
ions which can hereby be used for quantum computing \cite{lukinhemmer}.  
Along similar lines, the rare-earth-ion-doped inorganic crystal
proposal for quantum computing (REQC) was developed ~\cite{nicklas}. Instead
of excitation hopping, it uses a dipole blockade mechanism,
similar to the Rydberg blockade mechanism for neutral atoms \cite{rydberg},
which conditions gate operations on the interaction  
between a pair of ions exceeding
a certain threshold rather than having a precise value. The REQC proposal 
identifies qubit registers as entire ensembles of
collections of interacting ions, selected by their response to 
radiation at certain frequencies, and distributed over the spatial extent 
of the crystal. 

There is a
trade-off between the number of registers accepted by the spectroscopic selection 
and the precision we can request for their transition frequencies and
coupling strengths, and hence,
ultimately, between the scalability of the concept and the fidelity of the gate
operations. Gate operations are needed that can be carried out with
high fidelity, even if the physical parameters of the system are
allowed to vary, and following strong and convincing recommendations
by Jones \cite{jones}, we apply composite rotation
techniques from liquid and solid state NMR to the problem. 
Our method, which also involves quantum optical interference techniques, 
is described in detail for
the REQC scheme, but in the conclusion we shall comment on its possible
application to a number of other quantum computing proposals.

\section{Rare-earth-ion quantum computing scheme}
\begin{figure}
	\begin{center}
		\includegraphics[width=0.40\textwidth]{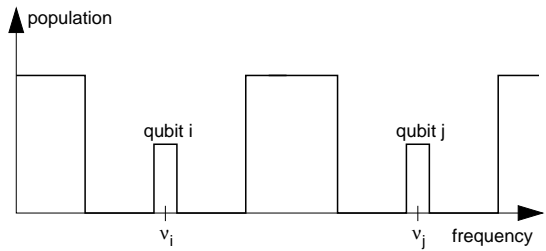}
	\end{center}
	\caption{A schematic diagram of a part of the inhomogeneously broadened
spectrum of excitation frequencies to the excited state \q{e} 
in rare-earth-ions in an inorganic
crystal. Two qubits \emph{i} and \emph{j} are defined by two different transition frequencies 
in the spectrally hole-burnt structure.}
	\label{fig:qubit}
\end{figure}

In the rare-earth ion scheme for 
quantum  computing, qubits are encoded in 
two ground hyperfine levels \q{0} and \q{1} of the dopant ions, which are 
only negligibly disturbed by the interaction with the 
crystal host. Excited states of the ions interact more strongly with the crystal,
and this gives rise to significant broadening of the optical transition between 
the qubit states and the excited state. The characteristic feature of the proposal
is the identification and initialization of the qubits and qubit registers in the crystal:  
The $i$th qubit is chosen as the ensemble of ions within the inhomogeneously broadened
line which have their absorption frequency 
at a chosen frequency $\nu_i$. Ions in a frequency window around the 
qubit channel are pumped to auxiliary states to prevent them from being excited by the 
near resonant laser light at the qubit frequencies, see Fig.\ \ref{fig:qubit}.

The excited state of an 
ion establishes a surrounding permanent dipole field, which 
shifts the transition frequencies in neighboring ions and which can 
hence control their interaction with a laser field.
Pairs of interacting qubits $(i,j)$ are hence identified, or rather selected, 
spectroscopically, by pumping away those ions which do not have their transition 
frequency shifted out of the frequency window around the channel frequency $\nu_{j(i)}$ 
by the dipole-field, when neighboring ions in the channel $i(j)$ are excited.
Physically, the crystal still contains all the ions, but around frequency
channels $\nu_{j(i)}$, the ions, populating the qubit levels, are found in close 
lying pairs, since all ions with insufficient 
excited state interaction have been deported to passive spectator levels. The ensemble of
ions in frequency channels $\nu_i$ and 
$\nu_j$ thus constitute pairs of mutually interacting ions.
Theoretical estimates \cite{nicklas} and preliminary experiments \cite{mattias}
suggest that the dipole energy shifts easily amount to several MHz, and the ions
can be excited with Rabi frequencies in the MHz range. The homogeneous linewidth is small
corresponding to excited
state lifetimes on the msec scale, whereas the qubit states have lifetimes of seconds with
coherence times in the msec range with possibilities for
increases up to several hundred msecs in the presence of magnetic fields \cite{annabel}.

\begin{figure}[tbp]
	\begin{center}
		\includegraphics[width=0.35\textwidth]{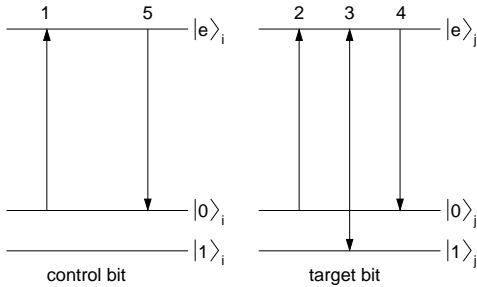}
	\end{center}
	\caption{A C-NOT operation with qubit $i$ as control and qubit $j\ $ as target bit. The operation makes 
use of the following sequence of $\pi$-pulses: 1. $\ket{0}_i \leftrightarrow \ket{e}_i$, 
2. $\ket{0}_j \leftrightarrow \ket{e}_j$, 
3. $\ket{1}_j \leftrightarrow \ket{e}_j$ 4. $\ket{0}_j \leftrightarrow \ket{e}_j$, and 5. $\ket{0}_i \leftrightarrow \ket{e}_i$.
See text for explanation.}
	\label{fig:cnot_concept}
\end{figure}

The implementation proposed in Ref.\ \cite{nicklas} for a C-NOT operation, with qubit $i$ 
as control bit and qubit $j$ as target bit, is shown in Fig.\ \ref{fig:cnot_concept}.
If qubit $i$ is initially in state \q{0}$_i$, it is excited by the first pulse, and the 
resulting frequency shift of
qubit $j$ makes the steps 2, 3 and 4 non-resonant, hence 
nothing happens, and qubit $i$ is finally returned to its initial state. 
If qubit $i$ is in state \q{1}$_i$, there is no frequency shift,  
and the resonant processes 2, 3 and 4, 
effectively exchange states \qq{0}{j} and \qq{1}{j}, as desired for the operation.
 
\section{High-fidelity $\pi$-pulses}
To have a sufficiently 
large number of active ions, and hence an appreciable number of quantum registers with, e.g., three or 
more coupled qubits, it is necessary to use ions with frequencies in a not too narrowly defined interval. 
The widths of the qubit channels, however, imply that the transitions 
1--5 in Fig.\ \ref{fig:cnot_concept} will not be resonant for all ions, 
and this seriously compromises the gate fidelity  in the system. One 
could apply very short pulses which are less sensitive to the resonance criterion, 
but they would 
excite ions outside the windows around the frequency channels, and they would not display the
required sensitivity 
to the frequency shift on ion $j$ due to the excitation 
of ion $i$ in the two-bit gate.  
In a recent paper \cite{cirac} it was proposed to use a combination of adiabatic
processes and coherent control theory to increase the tolerance to 
variations in physical  parameters in a quantum computing model, but 
very high fidelity was only achieved with very slow processes, which would 
be compromised by decoherence and decay in the REQC proposal. 
Instead, we shall present and 
compare the advantages of two different realizations
(composite rotation and complex 
hyperbolic secant pulse) of $\pi$-pulses, which rapidly transfer population completely
from one state to another in a two-level system, as long as the detuning of the 
transition lies within a 1 MHz wide  interval, but which does not transfer any
population if the
detuning exceeds just a few MHz. 
Composite pulses were, in fact,  already used in quantum computing experiments
with trapped ions \cite{blatt_nature}, not to correct for variations in
physical parameters, but to obtain $\pi$ and $4\pi$ Bloch sphere rotations
on two different transitions where the 
Rabi frequencies differed by a factor of $\sqrt{2}$.
We shall propose a combination of coherent population
trapping and $\pi$-pulse rotations as a means to make robust qubit rotations through arbitrary
angles and on arbitrary qubit states. 

\subsection{Composite pulses}
\begin{figure}[tbp]
	\begin{center}
		\includegraphics[width=0.45\textwidth]{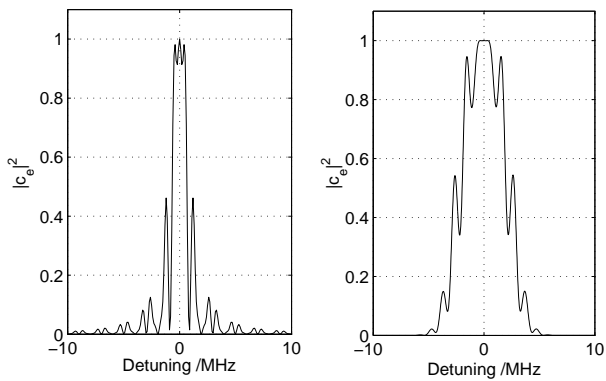}
	\end{center}
	\caption{Excitation by a composite pulse sequence $90_{90}180_{180}90_{90}$ of rectangular pulses  (left) and
a composite pulse sequence  $92.50_{96.98}192.00_{6.86}92.42_{96.23}$ of  Gaussian pulses (right).
The total duration of both sequences is 1.5 $\mu$s.}
	\label{fig:composite}
\end{figure}

Composite pulses can be assigned to two main classes, \emph{A} and \emph{B}, 
where pulses of type \emph{A} produce a fully compensated rotation of the system for all initial states, whereas 
pulses of type \emph{B} provide the compensated transformation only for  particular initial states. 
Levitt \cite{levitt} has tabulated several sequences of
rectangular pulses that provide net $90^{\circ}$ and $180^{\circ}$ Bloch sphere rotations 
of two-level systems and which are more tolerant to frequency detunings than a single
pulse.  Because of the high frequency components of the rectangular pulses, 
these pulses do not, however, satisfactorily exclude 
excitation of the surrounding ions, 
and we have instead studied composite pulses based on Gaussian field envelopes. On 
resonance, the accomplishments of a laser pulse is governed only by pulse area and phase, and hence it is natural
to replace a rectangular pulse sequence with a sequence of Gaussian pulses with the same areas and phases $\left(\theta_k,\varphi_k\right)$, i.e., with the time dependent Rabi frequencies
\begin{equation}
\Omega_R(t)=\sum_k \frac{\theta_{k}e^{-i\varphi_k}}{\sqrt{2\pi\sigma^2}} e^{-\left(t-t_k\right)^2/2\sigma^2}, \quad t\in [t_k-a,t_k+a].
\end{equation}
Composite rotations correct errors to high order, and there is a good portion of 
experience and magic 
in the design of good sequences. Merely replacing rectangular pulses with Gaussian 
pulses of equal pulse 
area and phase does therefore
not optimally preserve the  error compensating properties. 
The composite rotation with rectangular pulses, 
$90_{90}180_{0}90_{90}$ \cite{foot},
provides an error compensated $\pi$-pulse of type \emph{B}, and the left panel 
in \mbox{Fig.\ \ref{fig:composite}} illustrates its tolerance to detuning errors
on the 1 MHz scale, but also its erroneous excitation of ions more than 
5 MHz away from the channel frequency. 
Gaussian pulses with the same areas give a better performance without the off-resonant 
excitations, but we may do even better by slightly adjusting the pulse areas and phases. 
We have done this in a variational approach, assuming three pulses, and hence six free area and
phase parameters. For comparison, we choose the same duration of the rectangular
pulse and the Gaussian pulse sequences and a temporal cut-off $a=3.5 \sigma$, so that the 
pulses are truncated when the amplitude is $0.2\%$ of its maximum.
By insisting on perfect operation on resonance and minimum errors for a set of detunings, 
we have used a variational search to identify an optimum Gaussian pulse sequence with pulse 
areas and phases 
$92.50_{96.98}192.00_{6.86}92.42_{96.23}$.
Simulations of the excitation with this composite pulse are plotted in the right 
panel of Fig.\ \ref{fig:composite}. It has been been shown \cite{janus} that composite
pulses can satisfactorily deal with errors in coupling strengths,
e.g., due to spatial inhomogeneities of the field strength or variations of the transition
dipole moments in the ions,
but a gate that works for all input states was not identified in that work, 
and  no solution was presented for the 
important case of detuning errors in the REQC proposal.

\subsection{Complex hyperbolic secant pulse}
\begin{figure}[tbp]
  \begin{center}
		\includegraphics[width=0.45\textwidth]{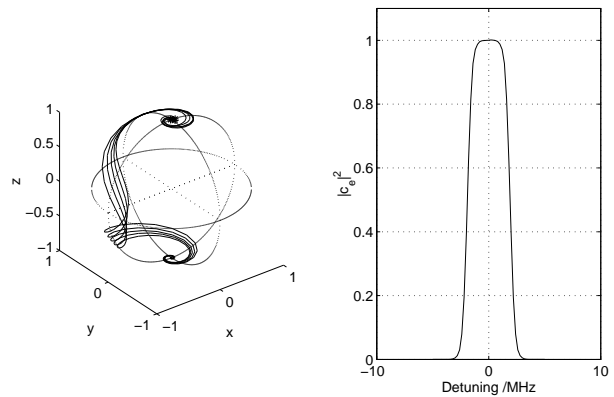}
	\end{center}
	\caption{Excitation by a complex hyperbolic secant pulse for different detunings. The left panel shows the time evolution on the Bloch sphere of ions with different detunings. 
The right panel shows the excited state population after the pulse as function of detuning. The parameters are $\mu=3$, $\Omega_0=2$ MHz and $\beta = 0.64$ MHz, and the duration of the pulse is \mbox{3 $\mu$s}.}
	\label{fig:secant}
\end{figure}

As shown by Silver et al.\ \cite{silver}, complex hyperbolic secant pulses offer an alternative process, of class \emph{B}, with the desired properties.
The complex Rabi frequency, $\Omega_R(t)$ of the complex hyperbolic secant pulse is given by
\begin{equation}
\label{eq:sech}
\Omega_R(t)=\Omega_0\left\{\mathrm{sech}\left[\beta\left(t-t_0\right)\right]\right\}^{1+i\mu},
\end{equation}
where $\mu$ is a real constant, $\Omega_0$ is the maximum Rabi frequency and 
$\beta$ is related to the FWHM of the pulse. This corresponds to a real sech pulse 
envelope in combination with a tanh frequency sweep around the carrier laser frequency
$\nu_L$:
\begin{eqnarray}
|\Omega_R(t)|& = & \Omega_0\mathrm{sech}\left(\beta\left(t-t_0\right)\right), \\ 
\nu(t) & = & \nu_L + \mu\beta\tanh\left(\beta\left(t-t_0\right)\right).
\end{eqnarray}
The excitation by such a pulse 
is shown in \mbox{Fig.\ \ref{fig:secant}} as a function of the detuning $(\nu_L-\nu_i)$. 
In addition to its favorable tolerance to detuning errors, the complex sech 
pulse is independent of pulse amplitude if only it exceeds a critical threshold. Its disadvantage compared to the composite Gaussian pulse is the double implementation time.

\section{Arbitrary rotations, compensation of phase shifts}
\subsection{Using dark states for arbitrary qubit rotations}
The pulses described are excellent for steps \mbox{1--2} and \mbox{4--5} in the C-NOT scheme in 
Fig.\ \ref{fig:cnot_concept}, since the initial states are 
explicitly known for all these processes due to the fact that the excited state is unpopulated when the operation starts. 
The initial state before step 3, however, is an unknown superposition state of \q{1}$_j$ and 
\q{e}$_j$, and a class~\emph{A} pulse is required to perform a perfect $\pi$-pulse 
on that state. 
Such composite pulses exist based upon square pulses, but they are difficult to build with
Gaussian pulses, and, we also find it difficult to make composite pulses for arbitrary
rotations, since the ones provided in the literature often involve a larger number of 
pulses of which some are rather short, and hence, they will interact with ions outside
the hole burnt structures.
Instead, we propose an alternative technique, which does not 
require class~\emph{A} pulses but nevertheless offers robust arbitrary rotations 
in the qubit space. To this purpose, we shall make use of the full three-level 
structure of the ions and
turn on fields with complex Rabi frequencies $\Omega_{R0}(t)e^{-i\varphi_0}$ and 
$\Omega_{R1}(t)e^{-i\varphi_1}$ that simultaneously
couple \qq{0}{j} and \qq{1}{j} to \qq{e}{j}. The system now has  two ground state superpositions, 
that will be respectively coupled and uncoupled to the excited state. If 
$\Omega_{R0}(t)=\Omega_{R1}(t)$ we 
can define a new orthonormal basis according to
\begin{equation}
\label{eq:base1}
\left\{ \begin{array}{ll}
\ket{\bar{0}} = \frac{1}{\sqrt{2}}\left(\ket{0}-e^{-i\phi}\ket{1}\right) \\
\ket{\bar{1}} = \frac{1}{\sqrt{2}}\left(\ket{0}+e^{-i\phi}\ket{1}\right)
\end{array} \right..
\end{equation}
If the relative phase of the two fields are chosen so that $\varphi_1-\varphi_0=\phi$, 
only \q{\bar{1}} couples to the 
excited state while \q{\bar{0}} is a dark state, unchanged by the interaction with the fields, 
and vice versa if $\varphi_1-\varphi_0=\phi+\pi$. 
Applying constant phase factors  $e^{-i\varphi_0},e^{-i\varphi_1}$ 
on the sequence of  Gaussian pulse or the 
complex sech pulse described above, we can therefore implement a robust $\pi$-pulse 
of class \emph{B}, i.e., from the south pole to the north pole on the Bloch sphere  
of states \q{e} and \q{\bar{1}}. If we drive the state back down again by a new $\pi$-pulse with a phase that is shifted by \mbox{$\pi+\theta$} in comparison to the first one, the net effect
becomes a robust phase shift on state \q{\bar{1}}, 
\begin{equation}
\label{eq:phase}
\ket{\bar{1}} \longrightarrow e^{i\theta}\ket{\bar{1}}.
\end{equation}
In the original qubit basis (\q{0},\q{1}), this procedure is equivalent to the unitary operation
\begin{equation}
U=e^{i\frac{\theta}{2}} \left( \begin{array}{cc}
\cos\frac{\theta}{2} & ie^{i\phi}\sin\frac{\theta}{2} \\
ie^{-i\phi}\sin\frac{\theta}{2} & \cos\frac{\theta}{2}
\end{array} \right) \label{eq:arb_matrix},
\end{equation}
i.e., a general unitary rotation in the qubit state basis.
A NOT operation is achieved if $\theta$ and $\phi$ are both set to $\pi$, but we note that
this approach offers any arbitrary qubit rotation, based exclusively on \mbox{class \emph{B}} 
$\pi$-pulses which, in turn, can be  implemented in a robust manner by
the pulses analyzed in Figs.\ \ref{fig:composite} and \ref{fig:secant}. 
A similar combination of coupled and uncoupled superposition states and the STIRAP
process was recently proposed \cite{kis}. 

\subsection{Phase compensation}
The Bloch sphere rotations introduce detuning and time dependent phases on the individual ions. 
A phase factor can be considered as global, and thus disregarded only if it appears in front of 
all qubit states. The two states involved in the above rotations are 
\q{\bar{1}} and \q{e} of the target bit, 
and one must necessarily arrange
for the same phase to be acquired by \q{\bar{0}}. We hence suggest to apply compensating pulses, 
which have no other effect than to introduce the same detuning dependent phase shift on that state,
that is a composite or complex sech $\pi$-pulse followed by the inverse 
$\pi$-pulse, i.e., with a phase shift of $\pi$ (but no $\theta$), 
on \mbox{$\ket{\bar{0}} \leftrightarrow \ket{e}$}. 
The choice of implementation and the duration of 
these pulses must be the same as those applied to 
\mbox{$\ket{\bar{1}} \leftrightarrow \ket{e}$} for the detuning dependent
phases to be accurately compensated. 

If the operation is a C-NOT operation, i.e. a two-bit operation, 
we will come across two different detuning dependent phase factors, 
one dependent on the detuning of the target ion and the other 
dependent on the detuning of the control ion. Since the detunings, and 
hence the phase factors, are unknown and not necessarily equal, 
we have to make sure that every qubit state acquires both phase factors 
in order to be able to consider the total unknown phase as global. 
The target qubit rotations, described above, only took place for the state vector
components with the control qubit in the state $|1\rangle$, and,
thus, conditioned on the control bit being in state \q{0}, we apply 
a new sequence of compensating 
pulses to the target states \q{\bar{0}} and \q{\bar{1}} so that all states 
\q{00}, \q{01}, \q{10} and \q{11} finally are equipped with 
the same detuning dependent phases. 

\begin{table}
	\begin{center}
		\begin{tabular}{|r|l|}
			\hline
			1. & $\pi$-pulse on $\ket{0}_i \leftrightarrow \ket{e}_i$ \\
			2. & $\pi$-pulse on $\ket{\bar{1}}_j \leftrightarrow \ket{e}_j$ \\
			3. & $\pi$-pulse on $\ket{\bar{1}}_j \leftrightarrow \ket{e}_j$ \\
			4. & $\pi$-pulse on $\ket{\bar{0}}_j \leftrightarrow \ket{e}_j$ \\
			5. & $\pi$-pulse on $\ket{\bar{0}}_j \leftrightarrow \ket{e}_j$ (inverse of pulse 4) \\
			6. & $\pi$-pulse on $\ket{0}_i \leftrightarrow \ket{e}_i$ (inverse of pulse 1) \\
		  7. & $\pi$-pulse on $\ket{1}_i \leftrightarrow \ket{e}_i$ \\
			8. & $\pi$-pulse on $\ket{\bar{1}}_j \leftrightarrow \ket{e}_j$ \\
			9. & $\pi$-pulse on $\ket{\bar{1}}_j \leftrightarrow \ket{e}_j$ (inverse of pulse 8)\\
			10. & $\pi$-pulse on $\ket{\bar{0}}_j \leftrightarrow \ket{e}_j$ \\
			11. & $\pi$-pulse on $\ket{\bar{0}}_j \leftrightarrow \ket{e}_j$ (inverse of pulse 10) \\
			12. & $\pi$-pulse on $\ket{1}_i \leftrightarrow \ket{e}_i$ (inverse of pulse 7) \\
			\hline
		\end{tabular}
	\end{center}
	\caption{Implementation scheme for a robust C-NOT operation. Pulses 4 and 5 compensate unknown phase factors, dependent on the detuning of the target bit, caused by pulses 2 and 3. Pulses 7--12 compensate phase factors dependent on the detuning of the control bit, caused by the sequence 1--6. All $\pi$-pulses should be implemented either
as composite pulse sequences or complex sech pulses to yield the desired tolerance to detuning variations.}
	\label{tab:scheme}
\end{table}

\section{Complete gate sequence, numerical results.}
\begin{figure}[tbp]
	\begin{center}
		\includegraphics[width=0.45\textwidth]{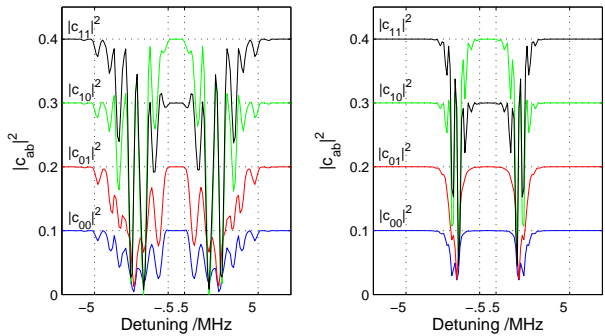}
	\end{center}
	\caption{Simulations of a C-NOT operation performed with optimized composite Gaussian pulses (left) and complex sech pulses (right). The initial qubit state was $\psi=\sqrt{0.1}\ket{00}+\sqrt{0.2}\ket{01}+\sqrt{0.3}\ket{10}+
\sqrt{0.4}\ket{11}$. The dotted vertical lines represent the edges of the qubit channels and of the spectral hole, respectively. The total duration of the operation 
is \mbox{18 $\mu$s} if all $\pi$-pulses are implemented with composite pulses and \mbox{36 $\mu$s} if they are
implemented with complex sech pulses.}
\label{fig:cnot_comp}
\end{figure}

The implementation scheme for a C-NOT operation  with qubit $i$ as control bit and qubit
$j$ as target bit is summarized in Table~\ref{tab:scheme}, and we draw the attention to the fact 
that 12 $\pi$-pulses
(which are all composite or complex sech pulses, but not of class \emph{A}) are applied. 
This scheme also suffices if the desired gate is a 
given rotation of the target bit, conditioned on the control bit; only the phase of
pulse 3 in the sequence should be adjusted accordingly.
Simulations according to this scheme have been performed by solving the equations of motion 
in the space spanned by the nine states \q{a_ib_j}, where $a,b=0,1,e$. An example
of the final state after action of the C-NOT operation on an initially entangled state
of the qubits is shown in \mbox{Fig.\ \ref{fig:cnot_comp}}.
For simplicity, 
the detuning parameter was assumed to be the same for the control and target ions
in the calculations leading to Fig.\ \ref{fig:cnot_comp}. The C-NOT operation effectively
exchanges the amplitudes and hence the populations on states 
$|10\rangle$ and $|11\rangle$, and this is precisely
what we observe for all detunings in the central 1 MHz wide frequency 
interval, whereas nothing happens to the state of ions more than 5 MHz away from the
exciting lasers. The intermediate frequency intervals are irrelevant, since they contain 
no ions, see \mbox{Fig.\ \ref{fig:qubit}}. Decay and decoherence become important on the
msec time scale, and it is hence important that the gates are fast on 
this time scale. The total duration of the operation is \mbox{18 $\mu$s} if all $\pi$-pulses are implemented with composite pulses and \mbox{36 $\mu$s} if they are implemented with complex sech pulses.
\begin{figure}
	\begin{center}
	\subfigure[Composite Gaussian pulses]
		{\includegraphics[width=0.45\textwidth]{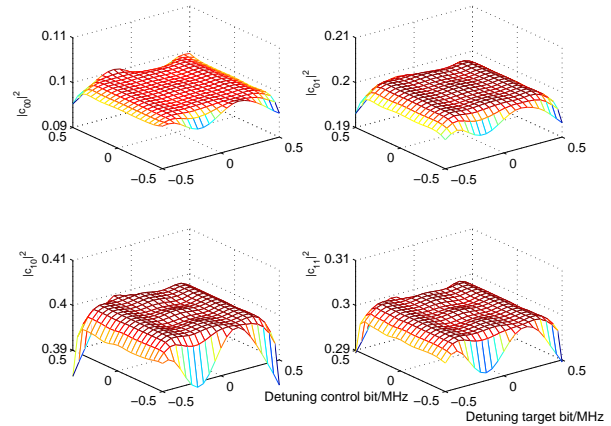}} \\
		\subfigure[Complex sech pulses]{\includegraphics[width=0.45\textwidth]{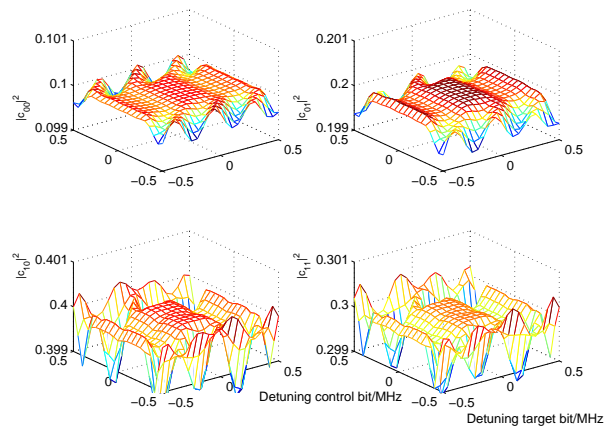}}
	\end{center}
	\caption{Simulations of a C-NOT operation where the control and target detunings are allowed to vary independently within the qubit frequency channel. The initial state was $\psi=\sqrt{0.1}\ket{00}+\sqrt{0.2}\ket{01}+\sqrt{0.3}\ket{10}+
\sqrt{0.4}\ket{11}$ and (a) is realized with composite Gaussian pulses while complex sech pulses are used in (b). Note the different scales.}
	\label{fig:cnot_comp_3d}
\end{figure}

In \mbox{Fig.\ \ref{fig:cnot_comp_3d}}, we show the results of our simulations when the control and target detunings
are allowed to both vary independently within the central 1 MHz wide frequency channel. The
four sub-plots show the populations, which are also presented in Fig.\ \ref{fig:cnot_comp}, and we see that
the deviations on the edges of the frequency interval are barely $10^{-3}$ when 
the 12 $\pi$-pulses are realized with complex sech pulses whereas 
implementation with composite Gaussian pulses results in deviations on
the order of  $10^{-2}$. The relative phases between the four states are accurate to 
the level of 1 degree for both kinds of pulses.

\section{Discussion}
We have shown that composite and complex sech pulses can be applied in combination 
with coherent population trapping to perform accurate quantum gate operations on ions in the REQC
quantum computing proposal. The composite Gaussian pulses do not reach the same high fidelity 
as the complex sech pulse. But, they are twice as fast and higher accuracies may be 
achieved with improved optimization.  
The combined tolerance to detuning errors on
one frequency scale and frequency selectivity on a single order of
magnitude larger scale are necessary to make quantum computing feasible at all
in the REQC proposal, and the first experiments to test our gates are in preparation.
Our theory also provide important ingredients for the further development
of the proposal, in particular it contributes significantly to the scalability to 
larger numbers of coupled qubits.

The achievements of the fault tolerant pulses 
are remarkable, and it is clear
that many quantum optics and quantum information tasks may benefit from
their introduction. Other proposals
exist for quantum computing where individual addressing of qubits
is made spectroscopically, e.g., of optically trapped neutral atoms 
in inhomogeneous magnetic fields \cite{lattice}, of trapped ions in
inhomogeneous magnetic \cite{wunderlich} and optical dipole fields \cite{staanum}.
Here, tolerance to variations in the detuning may be desirable as in the 
present work. Likewise, systems which use  spatial addressing may suffer from 
difficulties of focusing the fields on single qubits,
and here tolerance of the relevant qubit to variations around the maximum field  and
of the neighboring qubits to small variations around the desired vanishing field
can be accommodated by only slight modifications of the
techniques presented in this paper.

\section*{Acknowledgments}
Discussions with Stefan Kr\"oll and with Janus Wesenberg are gratefully acknowledged. 
This work was carried
out under the REQC and ESQUIRE projects supported by the IST-FET program of the EC.


\begin{thebibliography}{99}

\bibitem{lukinhemmer}M. D. Lukin and P. R. Hemmer, Phys. Rev. Lett. \textbf{84}, 2818 (2000).

\bibitem{nicklas}N. Ohlsson, R. K. Mohan, and S. Kr{\"o}ll, Opt. Commun. \textbf{201}, 71 (2002).

\bibitem{rydberg}D. Jaksch,
J. I. Cirac, P. Zoller, S. L. Rolston, R. C{\^o}t\'e, and M. D. Lukin 
Phys. Rev. Lett. \textbf{85}, 2208 (2000)

\bibitem{jones}J. A. Jones, quant-ph/0301019.

\bibitem{mattias}M. Nilsson, L. Rippe, N. Ohlsson, T. Christiansson, and S. Kr{\"o}ll, Phys. Scrip. \textbf{T102}, 178 (2002).

\bibitem{annabel} A. Alexander, {\it Investigation of qubit isolation in a rare-earth
quantum computer}. Honours thesis, Department of Physics, RSPSE, ANU (2000). 
The thesis is available at eprints.anu.edu.au/archive/00000761/.

\bibitem{cirac}J. J. Garc\'\i a-Ripoll and J. I. Cirac, Phys. Rev. Lett. \textbf{90}, 127902 (2003).

\bibitem{blatt_nature}S. Gulde \emph{et al.}, Nature \textbf{421}, 48-50 (2003).

\bibitem{levitt}M. H. Levitt, Prog. NMR Spectrosc. \textbf{18}, 61 (1986).

\bibitem{foot}$\theta_{\varphi}$ is used to denote a rotation through 
an angle $\theta$ about an axis in the $xy$-plane of the 
Bloch sphere at an angle $\varphi$ from the $x$-axis. 

\bibitem{janus}J. Wesenberg and K. M{\o}lmer, quant-ph/0301036.

\bibitem{silver}M. S. Silver, R. I. Joseph, and D. I. Hoult, Phys. Rev. A \textbf{31}, 2753 (1985).
\bibitem{kis} Z. Kis and F. Renzoni, quant-ph/0307208.

\bibitem{lattice} I. H. Deutsch and G. Brennen, Fortschr. Phys. {\bf 48}, 925 (2000).

\bibitem{wunderlich}F. Mintert and C. Wunderlich, Phys. Rev. Lett. \textbf{87}, 257904 (2001). 

\bibitem{staanum}P. Staanum and M. Drewsen, Phys. Rev. A \textbf{66}, 040302 (2002). 

\end{thebibliography}
\end{document}